# Mean-Reverting SABR Models:

# Closed-form Surfaces and Calibration for Equities

Vlad Perederiy*

December 2024, revised in October 2025

## Abstract

In this paper, we consider three stochastic-volatility models, each characterized by distinct dynamics of instantaneous volatility: (1) a CIR process for squared volatility (i.e., the classical Heston model); (2) a mean-reverting lognormal process for volatility; and (3) a CIR process for volatility. Previous research has provided semi-analytical approximations for these models in the form of simple (non-mean-reverting) SABR models, each suitably parameterized for a given expiry.

**First**, using a computer algebra system, we derive closed-form expressions for these semi-analytical approximations, under the assumption that all parameters remain constant (but without the constraint of constant expected volatility). Although the resulting formulas are considerably lengthier than those in simpler SABR models, they remain tractable and are easily implementable even in Excel.

**Second**, employing these closed-form expressions, we calibrate the three models to empirical volatility surfaces observed in EuroStoxx index options. The calibration is well-behaved and achieves excellent fits for the observed equity-volatility surfaces, with only five parameters per surface. Consequently, the approximate models offer a simpler, faster, and (numerically) more reliable alternative to the classical Heston model, or to more advanced models, which lack closed-form solutions and can be numerically challenging, particularly in less sophisticated implementation environments.

**Third**, we examine the stability and correlations of our parameter estimates. In this analysis, we identify certain issues with the models — one of which appears to stem from the sub-lognormal behavior of the actual equity-volatility process. Notably, the CIR-volatility model (3), as opposed to the CIR-variance Heston model (1), seems to best capture this behavior, and also results in more stable parameters.

---

* Perederiy Consulting (founder and senior consultant), PhD (Frankfurt/Oder University), FRM (GARP)
research@perederiy-consulting.de

## Acknowledgements

I would like to thank:

Patrick Hagan

(*World Bank, Gorilla Science*),

Jan-Philipp Hoffmann

(*Darmstadt University of Applied. Sciences / European University of Technology)*

Joerg Kienitz

(*University of Wuppertal, University of Cape Town*) and

Mirko Bono

(*DJE Investment*)

for their helpful discussions and insights.

## Contents

## 1. Introduction: stylized facts for equity volatilities

The returns of individual stocks and stock indices exhibit significant volatility clustering and mean reversion. Another typical feature (especially for indices) is the strong negative correlation between returns and changes in volatility. Figure 1 below illustrates these characteristics by showing the dynamics of the EuroStoxx index alongside the implied volatilities of index options of two different expiries.

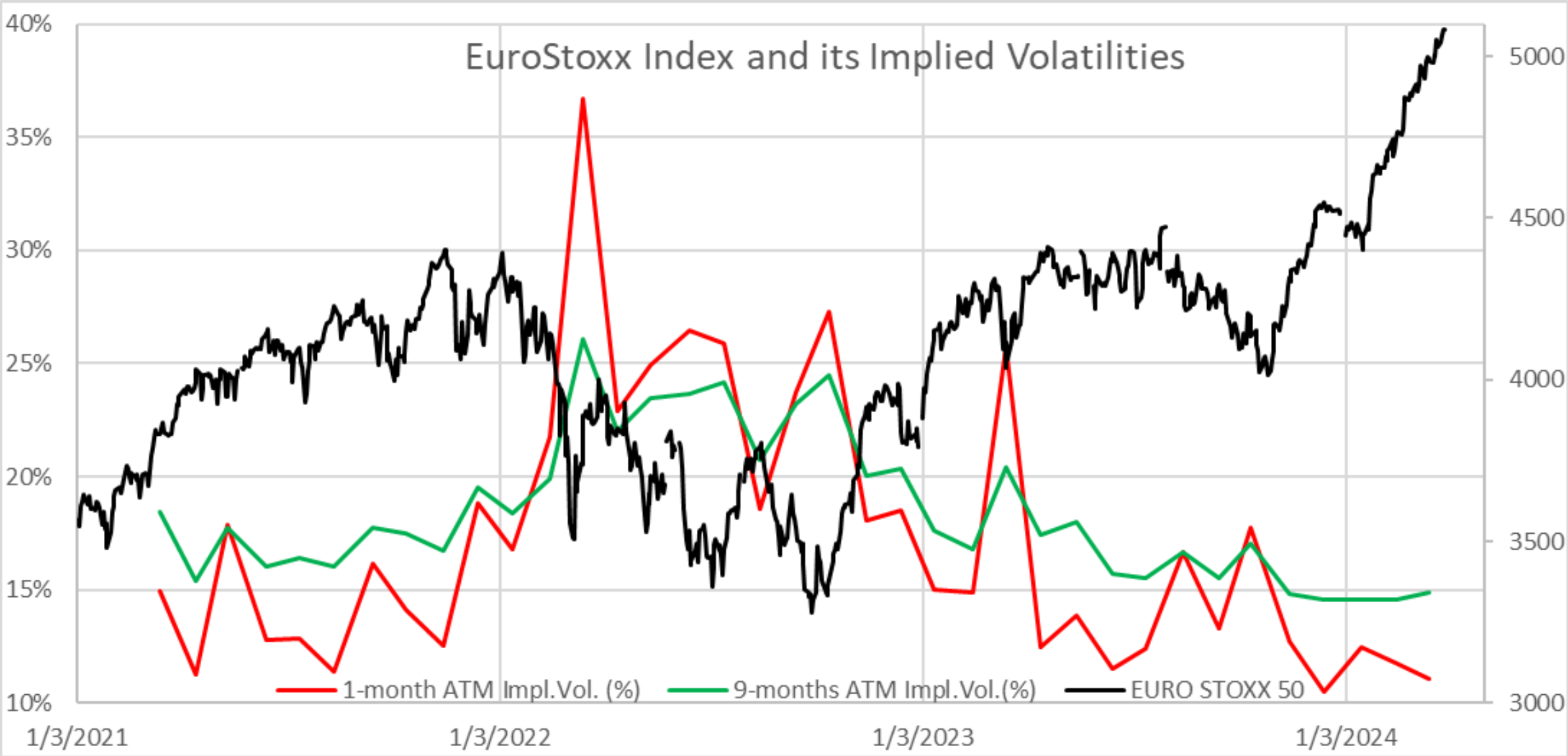

**Figure 1: Dynamics of EuroStoxx50 index and its implied volatilities**

These features are also reflected in equity derivatives, particularly in implied volatility surfaces. First, implied volatilities often exhibit a pronounced term structure: during periods of high volatility, shorter-term at-the-money (ATM) options typically show implied volatilities that are considerably higher than those of longer-term options (and vice versa). Second, because of the negative correlation, volatility smiles are generally skewed around the forward, resulting in so-called “volatility smirks.” Third, due to the mean reversion, the volatility smiles tend to flatten quickly for longer-term options.

Besides these features, typical trading and exchange standards mean that the relevant market quotes (option prices and/or implied volatilities) are directly observable only for a few strikes and maturities per underlying, with the availability and liquidity of the instruments varying over time. This makes the interpolation and extrapolation of observed market quotes—and their aggregation into volatility surfaces (across moneyness and expiry)—an important task in many applications, such as market risk management and valuation of complex derivatives.

Given the above characteristics, simple interpolation/extrapolation methods (e.g., parabola fitting of implied volatilities vs strike) are less suitable for equities than they are for, say, FX markets, where such approaches often suffice. Advanced fitting techniques—such as additional parametrization and strike/moneyness normalization (see, e.g., *Klassen, 2016)*—have been proposed. However, the resulting parameters tend to become increasingly mechanical (lacking clear economic interpretation) and still may not perform well in all situations. Figure 2 illustrates this challenge by showing volatility smiles in a rare scenario where the term structure of volatility is nearly flat. Even after moneyness normalization, the shape of the volatility smiles clearly differs here across maturities.

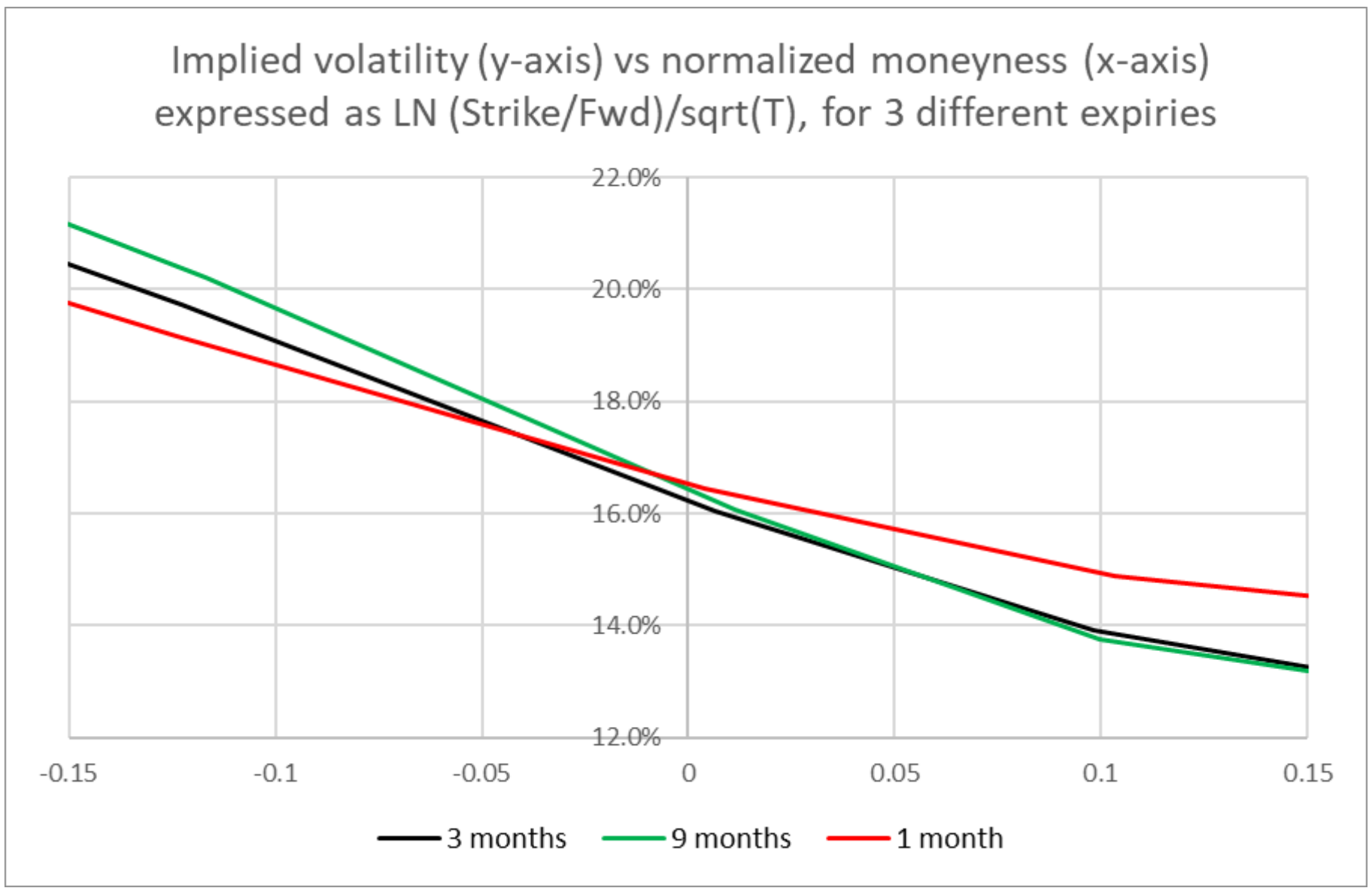

**Figure 2: Volatility smiles for EuroStoxx index options, an observation from August 2023 with nearly equal ATM volatilities**

Consequently, the industry standard for equities is typically to employ an underlying stochastic-volatility model that fits the observed implied volatilities for available strikes and maturities and predicts those for unobserved ones. The Heston model (*Heston*, *1993*) has long been the model of choice, as it captures the key characteristics of equity markets. More advanced models, such as local-stochastic volatility (LSV) models, have also been introduced.

A common drawback of these models is the lack of closed-form solutions. This necessitates advanced numerical methods for calibrating their parameters to empirical volatility surfaces, thereby limiting their use in many settings. In contrast, this paper discusses the application of recently developed mean-reverting SABR-based models, which capture the major characteristics of equity markets while still offering closed-form solutions.

## 2. Heston model

The Heston model can be specified as the following SDE system:

$$
\begin{aligned}
dF &= F\,\sqrt{V}\,dW_1 \\
d\mathrm{V} &= \lambda\,(\theta^2 - V)\,dt \;+\; \nu\,\sqrt{V} dW_2 \\
f &\equiv F_0 \\
\alpha &\equiv \sqrt{V_0} \\
dW_1\,dW_2 &= \rho\,dt
\end{aligned}
\tag{1}
$$

with following five parameters:

$\theta^2$: average/equilibrium variance (with $\theta \geq 0$ being the corresponding volatility)

$\alpha^2$: initial instantaneous variance (with $\alpha \geq 0$ being the corresponding volatility)

$\lambda$: speed of mean-reversion of variance, $\lambda \geq 0$

$\nu$: CIR volatility of variance, $\nu \geq 0$

$\rho$: correlation between the forward-price and variance processes, $-1 < \rho < 1$

Compared to the original *Heston (1993)* specification, we changed in (**1**) the parameter notations to those typically used in SABR models (which we discuss later), to make the models better comparable. For the same reason, we draw on the zero-drift process of the forward price $F$ instead of the (drifted) process of the spot price in the original Heston model. $f$ signifies the initial forward price and is normally not considered a parameter, as it can be inferred from the market data[1]. Also, we exclude the original Heston risk-premium parameter, as it is dispensable in the context of the market-implied measures.

Note that the variance V in Heston model follows a CIR process as specified in (**1**). The process for the volatility $A = \sqrt{V}$ in Heston is, upon applying the Ito's lemma:

$$
dA = \left( \frac{0.5\,\lambda\theta^2 - 0.125\,\nu^2}{A} - 0.5\,\lambda\,A \right) dt \;+\; \frac{1}{2}\,\nu\,dW_2 \tag{2}
$$

[1] The forward prices are typically not directly observable in equity markets. However, they can be readily inferred from spot prices by incorporating assumptions about dividends (where applicable) and interest rates. Additionally, forward prices can often be extracted from option prices using multiple call-put pairs through put-call parity.

which is a variant of normal-diffusion *Ornstein-Uhlenbeck* process, with a modified non-linear mean-reverting drift term[2].

For the above SDE framework (**1**), *Heston (1993)*, derived, using the techniques of characteristic functions and Fourier transforms, the seminal solution in terms of option prices. The solution is exact but semi-analytical, requiring numerical integration. Unfortunately, due to technical peculiarities in the solution, this often leads to numerical issues[3], and, in some cases, even results in negative calculated values for option prices.

This makes the Heston model and its solution less reliable and less suitable for simple environments, such as Excel, or for unsupervised settings like Monte Carlo simulations, even when simply calculating option prices from its five given parameters. The issues aggravate when the parameters are unknown and have yet to be calibrated to best fit some market data (such as a volatility surface).

Apart from the problematic numerical estimation, another issue with the Heston model is the so-called Feller condition:

$$2\,\lambda\,\theta^2 > \nu^2 \tag{3}$$

If (**3**) is violated, the variance process in (**1**) might, in theory, degenerate. However, market fits often do not satisfy this condition. This makes extrapolating volatility surfaces using a Heston-fit model quite risky, as it may produce implausible degenerative implied volatilities for the regions (strikes and expiries) situated far from the data points used for the surface calibration.

Finally, as already mentioned, the Heston (1993) solution is in the form of option prices. When calibrating to volatility surfaces, these prices have yet to be converted to Black-implied volatilities, resulting in additional computational expense and sometimes also in additional numerical-precision issues.

## 3. Standard SABR model

The standard SABR model is based on the following SDE system:

$$dF = F^{\beta}\,\mathrm{A}\,dW_1 \tag{4}$$

---

[2] Note also that the equilibrium volatility $A_{eq}$ in Heston is calculated as follows: $A_{eq} = \sqrt{\theta^2 - 0.25\,\nu^2/\,\lambda}$.

[3] The integrals involved are over osculating functions using logarithms of complex numbers, and are subject to branching/discontinuity issues. See *Cui et al. (2017)* for details and possible remedies.

$$dA = \nu\, A\, dW_2$$

$$f \equiv F_0$$

$$\alpha \equiv A_0$$

$$dW_1 dW_2 = \rho\, dt$$

with the parameters as follows:

$\beta$: so-called backbone parameter, $0 \leq \beta \leq 1$

$\alpha$: initial instantaneous volatility, $\alpha \geq 0$

$\nu$: lognormal volatility of volatility, $\nu \geq 0$

$\rho$: correlation between the forward-price and volatility processes, $-1 < \rho < 1$

$f$ is the initial forward price inferred from the market data as before.

The backbone parameter $\beta$ is normally set beforehand, using market standards/beliefs, effectively making the standard SABR a model with **three-parameters** $\alpha, \nu, \rho$.

Using perturbation techniques, *Hagan et al. (2002)* derived a simple approximate closed-form solution for the SDE system **(4)**. The solution is specified in terms of implied volatilities of a European option, either Bachelier (normal) or Black (lognormal) ones. Specifically, for Black implied volatilities, which are the standard in equity market quotations, following approximation can be used:

$$\sigma_B = \frac{z}{x(z)}\ \frac{\alpha}{f_{av}^{1-\beta}}\ \frac{1 + \left(\frac{(1-\beta)^2}{24}\frac{\alpha^2}{f_{av}^{2-2\beta}} + \frac{\beta}{4}\rho\nu\frac{\alpha}{f_{av}^{1-\beta}} + \frac{2-3\rho^2}{24}\nu^2\right)T_{ex}}{1 + \frac{(1-\beta)^2}{24}\log^2\frac{f}{K} + \frac{(1-\beta)^4}{1920}\log^4\frac{f}{K}} \tag{5}$$

with:

$$f_{av} = \sqrt{fK}$$

$$z = \frac{\nu}{\alpha} f_{av}^{1-\beta} \log\frac{f}{K}$$

$$x(z) = \log\left(\frac{\sqrt{1-2\rho z + z^2} + z - \rho}{1-\rho}\right)$$

where $K$ signifies the strike of the option and $T_{ex}$ its expiry (maturity).

In the special case $K = f$ (forward-ATM options) we have:

$$\frac{z}{x(z)} \to 1$$

$$\sigma_{B,ATM} = \frac{\alpha \left\{1 + \left(\frac{(1-\beta)^2}{24}\frac{\alpha^2}{f^{2-2\beta}} + \frac{1}{4}\frac{\rho\beta\nu\alpha}{f^{1-\beta}} + \frac{2-3\rho^2}{24}\nu^2\right)T_{ex}\right\}}{f^{1-\beta}} \quad (6)$$

The above original *Hagan et al. (2002)* approximations are probably still most widely used, due to their simplicity. Several other closed-form variants, with improved approximative precision, have been proposed later. All of them remain of the same order *O(2)*, however. For an example, see e.g. *Hagan et al. (2016).*

For equities, the special case of so-called lognormal-SABR with $\beta = 1$ is the most natural, as it corresponds to the widely accepted assumption of (approximative) lognormality in asset prices. In this case, the implied Black volatilities in (**5**) and (**6**) conveniently reduce to:

$$\sigma_B = \frac{z}{x(z)}\alpha \left\{1 + \left(\frac{1}{4}\rho\nu\alpha + \frac{2-3\rho^2}{24}\nu^2\right)T_{ex}\right\} \quad (7)$$

with:

$$z = \frac{\nu}{\alpha}\log\frac{f}{K}$$

$$x(z) = \log\left(\frac{\sqrt{1-2\rho z + z^2} + z - \rho}{1-\rho}\right)$$

and $\frac{z}{x(z)} \equiv 1$ for ATM options (where $K = f$).

As stated in *Hagan et al. (2002)*, (**7**) actually happens to provide a more exact *O(4)* approximation for the special case of lognormal-SABR model.

In principle, the standard SABR model (**4**), with $\beta = 1$ and approximative closed-form solution such as in (**7**), can be easily calibrated to a given equity volatility smile of a certain maturity. However, a SABR calibration to a whole equity volatility surface (across the additional expiry dimension) would be highly problematic. The model only partially reflects the features of equity markets (see Section 1): whereas the negative return-volatility correlation will be adequately captured by its $\rho$ parameter, the term structure and mean-reverting in volatility are not accounted for in the standard SABR model.

This leads to the situation where the standard SABR model, despite its concise closed-form solution and its huge popularity in certain other markets (in particular, IR), is rarely used by equity practitioners. In principle, the standard SABR model can still be used for equity surface calibration if estimated separately for different maturities. This would typically lead to highly maturity-dependent parameter estimates in (**4**), in particular:

- $\nu$ would strongly diminish with increasing maturity/expiry, as a compensation for the unrestricted volatility diffusion in the standard SABR model
- $\alpha$ would often differ considerably across expiries, also reflecting the mean reversion

The separately estimated standard-SABR models could then be combined into surfaces using some additional heuristic techniques, such as interpolating parameter estimates across expiries. However, a more consistent model that explicitly accounts for mean reversion and the term structure in volatility should clearly be the preferred solution for equities

## 4. Heston-SABR and its semi-analytical solution (hSABR)

Given the computational challenges with the exact solution of the Heston model as depicted in Section 2, its simpler closed-form approximations will be preferable in many settings.

*Hagan et al. (2018)* considered the classical Heston model (**1**), with its parameters as given, and derived an approximation using the advanced techniques of effective forward equations, singular perturbation and effective media theory. The approximation in the form of a standard SABR model (**4**) with suitable three parameters which depend on the five Heston parameters and on the option's expiry. Given these standard-SABR parameters, the already mentioned closed-form approximations (**7**) can be used to infer the implied Black volatilities.

The solution in *Hagan et al. (2018)* is a multi-step procedure resulting in a semi-analytical solution through nested integrals. We re-iterate and somewhat simplify this solution (which we call *hSABR*) below.

Starting point is the function $V(T)$ of the expected value for the variance at a future time point $T$, which can be derived from (**1**) as follows:

$$V(T) = E(V(T) \mid V_0 = \alpha^2) = \alpha^2 e^{-\lambda T} + \theta^2 \left(1 - e^{-\lambda T}\right) \tag{8}$$

Then, the intermediate integral functions are defined as follows:

$$I_2(T) = \rho\nu \int_0^T V(T_1) \int_{T_1}^T e^{-\lambda(T_2-T_1)}\, dT_2\; dT_1 \tag{9}$$

$$I_4(T) = \rho^2\nu^2 \int_0^T V(T_1)\, e^{-\lambda(T-T_1)}\, (T-T_1)\, dT_1$$

$$D(T) = \int_T^{T_{ex}} e^{-\lambda(T_1-T)}\; dT_1$$

Note that in (**9**) we used the original numbering/denominations for the integrals from *Hagan et al. (2018)* but skipped some intermediate integrals not used in the final results. Also, we simplified the integrals by assuming constant parameters (as in the original Heston model), thus refraining from their time-dependent generalizations in *Hagan et al. (2018)*[4].

The next step in *Hagan et al. (2018)* is the calculation of the so-called **effective coefficients** which are defined as follows for an option's expiry $T_{ex}$:

$$\tau_{ex} = \int_{T_1}^{T_{ex}} V(T_1)\; dT_1 \tag{10}$$

$$\bar{b} = \frac{I_2(T_{ex})}{\tau_{ex}^2}$$

$$\bar{c} = \frac{3}{\tau_{ex}^3} c_{int} - 3\bar{b}^2 \quad \text{with } c_{int} = \int_0^{T_{ex}} \frac{\nu^2}{4} V(T)\, D^2(T) + I_4(T)\; dT$$

Finally, the three standard SABR coefficients are calculated as follows from the effective coefficients:

$$\alpha_{std} = \sqrt{\frac{\tau_{ex}}{T_{ex}}}\;\; \mathrm{e}^{-\frac{\bar{c}}{4}\tau_{ex}} \tag{11}$$

$$\rho_{std} = \frac{\bar{b}}{\sqrt{\bar{c}}}$$

[4] Also, compared to *Hagan et al. (2018*), we omitted the σ parameter which is redundant in the case of constant parameters.

$$\nu_{std} = \sqrt{\frac{\tau_{ex}}{T_{ex}}\bar{c}}$$

*Hagan et al. (2018)* showed that the standard SABR model (**4**) with the three parameters $\alpha$, $\rho$ and $\nu$ calculated as in (**11**) approximates the original five-parameter *Heston* model (**1**) well, up to the order *O(2)*, for a given expiry $T_{ex}$. This latter dependency on the expiry accounts for the mean-reverting volatility in the Heston model.

## 5. Mean-reverting SABR and its semi-analytical solution (mrSABR)

The mean-reverting SABR model as in *Hagan et al. (2020)* is no more based directly on the Heston model. Instead, it adds a mean-reverting term directly to the standard SABR model (**4**), with the SDEs now becoming:

$$\begin{aligned} dF &= F^{\beta}\,\mathrm{A}\,dW_1 \\ d\mathrm{A} &= \lambda\,(\theta - A)\,dt + \nu\,A\,dW_2 \\ f &\equiv F_0 \\ \alpha &\equiv A_0 \\ dW_1 dW_2 &= \rho dt \end{aligned} \tag{12}$$

with following five parameters:

$\theta$: average/equilibrium volatility, $\theta \geq 0$

$\alpha$: initial instantaneous volatility, $\alpha \geq 0$

$\lambda$: speed of mean-reversion of volatility, $\lambda \geq 0$

$\nu$: lognormal volatility of volatility, $\nu \geq 0$

$\rho$: correlation between the forward-price and volatility processes, $-1 < \rho < 1$

and the backbone parameter $\beta$ (which can be set to 1 for equities, as explained above).

Note the similarity between (**12**) and the classical Heston model (**1**), though they are not identical. Most importantly, as can be seen from (**2**), the diffusion process for volatility is nearly mean-reverting normal under Heston, whereas it is mean-reverting lognormal in (**12**). Thus, even from a theoretical perspective, we can expect some differences in the behavior of the two models. Besides, the volatility process can be shown to remain non-degenerating for (**12**) if the following condition is satisfied:

$$\lambda > \frac{1}{2}\nu^2 \tag{13}$$

Note the difference to the Feller condition (**3**) in Heston.

Proceeding analogously to *hSABR* (see Section 4), *Hagan et al. (2020)* derive a semi-analytical *O(2)* approximation (which we refer to as *mrSABR*) for the model specification in (**12**) in terms of a standard three-parameter SABR model. We re-iterate this solution below.

The starting point is the function $\alpha(T)$ as expected value of future volatility at a time point $T$:

$$\alpha(T) = E(A(T) \mid A_0 = \alpha) = \alpha e^{-\lambda T} + \theta\left(1 - e^{-\lambda T}\right) = \theta + (\alpha - \theta)\, e^{-\lambda T} \tag{14}$$

Then, the five intermediate integral functions $I_1(T)$-$I_5(T)$ are defined as follows:

$$I_1(T) = \rho\nu \int_0^T \alpha^2(T_1)\, e^{-\lambda(T-T_1)}\, dT_1 \tag{15}$$

$$I_2(T) = \nu^2 \int_0^T \alpha^2(T_1)\, e^{-\lambda(T-T_1)} \int_{T_1}^T \alpha(T_2)\, e^{-\lambda(T_2-T_1)}\, dT_2\ dT_1$$

$$I_3(T) = \rho\nu \int_0^T \alpha^2(T_1) \int_{T_1}^T \alpha(T_2)\, e^{-\lambda(T_2-T_1)}\, dT_2\ dT_1$$

$$I_4(T) = \rho^2\nu^2 \int_0^T \alpha^2(T_1)\, e^{-\lambda(T-T_1)} \int_{T_1}^T \alpha(T_2)\ dT_2\ dT_1$$

$$I_5(T) = \nu^2 \int_0^T \alpha^2(T_1)\, e^{-2\lambda(T-T_1)}\, dT_1$$

Again, when specifying (**15**) we made some simplifications compared to the original *Hagan et al. (2020)* paper (see Section 4 for details). Analogous to *hSABR*, the next step in the *mrSABR* approach is calculating the effective coefficients for the option's expiry $T_{ex}$:

$$\tau_{ex} = \int_0^{T_{ex}} \alpha^2(T)\ dT \tag{16}$$

$$\bar{b} = \frac{2}{\tau_{ex}^2} I_3(T_{ex})$$

$$\bar{c} = \frac{3}{\tau_{ex}^3} c_{int} - 3\bar{b}^2 \quad \text{with } c_{int} = \int_0^{T_{ex}} 2\alpha(T)\, I_2(T) + I_1^2(T) + 4\alpha(T) I_4(T)\, dT$$

$$G_{int} = \int_0^{T_{ex}} I_5(T) \quad dT$$

Finally, the three so-called standard SABR coefficients are calculated from the effective coefficients:

$$\alpha_{std} = \sqrt{\frac{\tau_{ex}}{T_{ex}}} \quad e^{-\frac{\bar{c}}{4}\tau_{ex} + \frac{G_{int}}{2\,\tau_{ex}}} \tag{17}$$

$$\rho_{std} = \frac{\bar{b}}{\sqrt{\bar{c}}}$$

$$\nu_{std} = \sqrt{\frac{\tau_{ex}}{T_{ex}} \bar{c}}$$

Again, *Hagan et al. (2020)* showed that the standard SABR model (**4**) with the three parameters $\alpha$, $\rho$ and $\nu$ calculated as in (**17**) approximates the five-parameter model specified by (**12**) well, up to the order *O(2)*, for a given expiry $T_{ex}$.

Intuitively, despite formally the same approximating order *O(2)*, the goodness of the *mrSABR* approximation might be better than in the case of *hSABR*, as the approximating model (standard-SABR model (**4**)) seems somewhat closer to the former target model (mean-reverting SABR model (**12**)) than to the latter target model (Heston model (**1**)).

## 6. Mean-reverting ZABR and its semi-analytical solution (mrZABR)

*Felpel et al. (2020)* investigated the model with a generalization of the diffusion process of volatility:

$$dF = F^{\beta}\, \mathrm{A}\, dW_1 \tag{18}$$

$$d\mathrm{A} = \lambda\,(\theta - A)\, dt + v(A)\, dW_2$$

$$f \equiv F_0$$

$$\alpha \equiv A_0$$

$$dW_1 dW_2 = \rho dt$$

with some function $v(A)$. Note that for $v(A) = v\, A$, the model reduces to *mrSABR* as in (**12**). *Felpel et al. (2020)* provided the derivation of a semi-analytical solution analogous to *mrSABR*. We first reiterate this solution, focusing on the differences to *mrSABR.*

First, with an additionally defined function $\psi(A) = v(A)\, A$ and the expected volatility function $\alpha(T)$ remaining as in **(14)**, the intermediate integral functions are defined as follows:

$$I_1(T) = \rho \int_0^T \psi\big(\alpha(T_1)\big)\; e^{-\lambda(T-T_1)}\, dT_1 \tag{19}$$

$$I_2(T) = \int_0^T \Big(v\big(\alpha(T_1)\big)\Big)^2\; e^{-2\lambda(T-T_1)} \int_{T_1}^T \alpha(T_2)\, e^{\lambda(T-T_2)}\, dT_2\;\, dT_1$$

$$I_3(T) = \rho \int_0^T \psi\big(\alpha(T_1)\big)\; e^{-\lambda(T-T_1)} \int_{T_1}^T \alpha(T_2) e^{\lambda(T-T_2)}\, dT_2\;\, dT_1$$

$$I_4(T) = \frac{1}{2}\rho^2 \int_0^T \psi\big(\alpha(T_1)\big)\; e^{-\lambda(T-T_1)} \int_{T_1}^T \psi'\big(\alpha(T_2)\big)\, dT_2\, dT_1$$

$$I_5(T) = \int_0^T \Big(v\big(\alpha(T_1)\big)\Big)^2\; e^{-2\lambda(T-T_1)}\, dT_1$$

Note that, compared to the original *Felpel et al. (2020)* paper, we slightly redefined $\psi$, rearranged terms, and also multiplied the corresponding integrals $I_2(T)$ and $I_4(T)$ by a factor of ½, for the sake of simplification and better comparability with *mrSABR.*

Then, the effective coefficients and standard SABR coefficients can be calculated analogously to *mrSABR* as in (**16**) and (**17**). *Felpel et al. (2020)* showed that the resulting standard SABR model approximates the generalized model specified in (**18**) well, up to the order *O(2)*, for a given expiry $T_{ex}$.

The so-called **mean-reverting ZABR** model (*mrZABR*) is a special case of the above model with:

$$v(A) = v\, \mathrm{A}^{\gamma} \tag{20}$$

$$\psi(A) = \nu\, A^{\gamma+1}$$

$$\psi'(A) = (\gamma + 1)\, \nu\, A^{\gamma}$$

Here, the volatility follows a mean-reverting CEV process. Not surprisingly, for $\gamma = 1$ the integrals in (**19**) can be shown to be equivalent to the *mrSABR* integrals (**15**), and the model reduces to *mrSABR*. For $\gamma = 0$, volatility diffusion becomes similar to that for Heston in (**2**). Setting $\gamma = 0.5$ places the model between *hSABR* and *mrSABR*. Here, a CIR process is assumed for the volatility (unlike for the variance as in Heston), with the Feller condition now becoming:

$$2\,\lambda\,\theta > \nu^2 \tag{21}$$

We refer to this latter *mrZABR* specification with $\boldsymbol{\gamma = 0.5}$ as ***CIR-ZABR***.

## 7. Closed-form solutions: special case of constant expected volatility

Apart from the semi-analytical solutions in terms of nested integrals, as in (**9**)-(**10**) for *hSABR* and (**15**)-(**16**) for *mrSABR*, *Hagan et al. (2018)* and *Hagan et al. (2020)* derive explicit closed-form formulas for all-constant parameters with the additional restriction $\alpha = \theta = \sigma$, i.e. with both initial instantaneous volatility and long-term volatility equal to some common value $\sigma$. In this special case, the expected variance $V(T)$ in (**8**) in *hSABR* becomes constant and equals $\sigma^2$, and the expected volatility $\alpha(T)$ in (**14**) in *mrSABR* equals $\sigma$, for all future time points $T$. This considerably simplifies the integrals mentioned.

The semi-analytical expressions for the effective *hSABR* coefficients in (**9**)-(**10**) reduce then, in this special case, to[5]:

$$\tau_{ex} = \sigma^2 T_{ex} \qquad \bar{b} = \frac{\rho\nu}{\sigma^2}\,\frac{\lambda T_{ex} - 1 + e^{-\lambda T_{ex}}}{\lambda^2\, T_{ex}^2} \tag{22}$$

$$\bar{c} = \frac{3\nu^2}{\sigma^4}\,\frac{1 + 2\lambda T_{ex} - \left(2 - e^{-\lambda T_{ex}}\right)^2}{8\,\lambda^3\, T_{ex}^3} + 3\,\frac{\rho^2\nu^2}{\sigma^4}\,\frac{\lambda^2\, T_{ex}^2 e^{-\lambda T_{ex}} - \left(1 - e^{-\lambda T_{ex}}\right)^2}{\lambda^4\, T_{ex}^4}$$

For *mrSABR*, the corresponding effective coefficients in (**15**)-(**16**) become, in this special case:

---

[5] Note that the corresponding formula in *Hagan et al. (2018)* differs in the power of $\sigma$. This is due to a different definition of this parameter in *Hagan et al. (2018)* where it is an additional scaling coefficient $\sigma$ in the forward process $F = \sigma\, F\, \sqrt{V}\, dW_1$, with the $\alpha = \theta$ restriction being implemented there via $\alpha = \theta = 1$.

$$\tau_{ex} = \sigma^2 T_{ex} \qquad \bar{b} = \frac{2\rho\nu}{\sigma} \frac{\lambda T_{ex} - 1 + e^{-\lambda T_{ex}}}{\lambda^2\, T_{ex}^2} \tag{23}$$

$$\bar{c} = \frac{3\nu^2}{\sigma^2}(1+\rho^2)\frac{1 + 2\lambda T_{ex} - \left(2 - e^{-\lambda T_{ex}}\right)^2}{2\,\lambda^3\, T_{ex}^3} + 12\frac{\rho^2\nu^2}{\sigma^2}\frac{\lambda^2\, T_{ex}^2 e^{-\lambda T_{ex}} - \left(1 - e^{-\lambda T_{ex}}\right)^2}{\lambda^4\, T_{ex}^4}$$

$$G_{int} = \frac{\nu^2\sigma^2\left(2\lambda T_{ex} - 1 + e^{-2\lambda T_{ex}}\right)}{4\,\lambda^2}$$

Lastly, *Felpel et al. (2020)* derived the corresponding effective coefficients for the *mrZABR* model in the special case $\alpha = \theta = \sigma$ as[6]:

$$\bar{b} = 2\rho\nu\sigma^{\gamma-2}\, \frac{\lambda T_{ex} - 1 + e^{-\lambda T_{ex}}}{\lambda^2\, T_{ex}^2} \tag{24}$$

$$\bar{c} = \frac{3\,(1+\rho^2)\nu^2\sigma^{2(\gamma-1)-2}}{2\,\lambda^3\, T_{ex}^3}\left(2\lambda T_{ex} + 4e^{-\lambda T_{ex}} - 3 - e^{-2\lambda T_{ex}}\right)$$
$$+6\frac{(1+\gamma)\rho^2\nu^2\sigma^{2(\gamma-1)-2}}{\lambda^3\, T_{ex}^3}\left(\lambda T_{ex} + 2e^{-\lambda T_{ex}} - 2 + \lambda T_{ex} e^{-\lambda T_{ex}}\right)$$
$$-12\rho^2\nu^2\sigma^{2(\gamma-1)-2}\left(\frac{\lambda T_{ex} - 1 + e^{-\lambda T_{ex}}}{\lambda^4\, T_{ex}^4}\right)^2$$

As expected, the *mrZABR* expressions in (**24**) reduce, for $\gamma = 1$, to the *mrSABR* ones in (**23**).

The standardized SABR coefficients can then be calculated, as before, via the simple logic in (**11**) for hSABR or (**17**) for *mrSABR* and *mrZABR*.

## 8. Derivation of closed-form solutions without restrictions

Although useful as a simplification, the restriction $\alpha = \theta$ when deriving the closed-form expressions in the previous research (see Section 7) makes the model less suitable for the calibration to volatility surfaces in equity markets, because of the market specifics mentioned above (see Section 1). We now return to the semi-analytical solutions as in the Sections 4, 5 and 6 above, **refrain from the restriction $\boldsymbol{\alpha = \theta}$** and show below that the solutions can still be reduced to closed-form expressions.

[6] Note that the definitions of $\bar{b}$ and $\bar{c}$ in *Felpel et al. (2020)* differ slightly from the *mrSABR* definitions in *Hagan et al. (2020)*, which we also follow. To reconcile the definitions, we divided the original expressions from *Felpel et al. (2020)* by a factor of $\sigma$ for $\bar{b}$ and by a factor of $\sigma^2$ for $\bar{c}$, leading to (**24**). Additionally, we corrected a typo (missing power of 2 in $\bar{c}$) in *Felpel et al. (2020).*

At the first sight, the derivation of analytical expressions for the effective coefficients in (**10**) in the case of *hSABR*, or (**16**) in the case of *mrSABR*, from the original parameters $(\alpha, \theta, \lambda, \rho, \nu)$, via the nested integral functions (**9**) and (**15**), looks quite challenging, especially in the case of $\bar{c}$ for *mrSABR*. However, all of them do result in closed-form expressions without any further approximations. The derivation itself is quite straightforward, via common integration rules, but very tedious. Especially in the case of $\bar{c}$ for *mrSABR*, the derivation is barely doable without computer assistance. We retorted to the **symbolic integration** routine from the python *sympy* package, and obtained the results as follows. Each of the effective coefficients $\tau_{ex}$, $G_{int}$, $\bar{b}$, $\bar{c}$, both for *hSABR* and *mrSABR*, eventually reduces to a simple **ratio of two multivariate polynomials** over the following 7 variables:

- The 5 original parameters $\alpha, \theta, \lambda, \rho, \nu$

- Maturity/expiry $T_{ex}$

- The exponential term $E \equiv e^{\lambda T_{ex}}$

Appendix A and B to this paper contains the resulting closed-form expressions for *hSABR* and *mrSABR* in the Excel-formula format, with the polynomials in clustered/factorized form, as optimized by *sympy*. These expressions are easily implementable in Excel[7]. When expanded, the polynomials get very lengthy (e.g. over 200 members in the nominator of $\bar{c}$ for *mrSABR*) and run up to the degree 8 of the variables mentioned.

For the more general *mrZABR* model, the integrals (**19**) with the function specification (**20**) cannot be reduced to closed-form expressions in terms of elementary functions. The problematic parts are the integrand terms of the type $\left(\alpha(x)\right)^g$ with $\alpha(x) = \theta + (\alpha - \theta)\, e^{-\lambda x}$, and $g = \gamma$ or $g = \gamma + 1$. They do not have analytical antiderivatives with respect to $x$ in elementary functions, apart from special cases (such as $\gamma = 1$).

[7] An implementation via direct (cell-inserted) Excel formulas might be easier than via VBA, as each Excel cell formula allows for up to 8192 characters, whereas a line of VBA code only allows for around 1023 symbols at a time (which is surpassed e.g. in case of $\bar{c}$ for *mrSABR*).

To deal with this problem, we used further approximations of all integrand parts involving such terms, prior to the analytical integration. In particular, we retorted to the method of binomial expansion, which is generally specified as follows:

$$(1+y)^g \approx \sum_{n=0}^{\infty} \binom{g}{n} y^g = 1 + gy + \frac{g(g-1)}{2!} y^2 + \frac{g(g-1)(g-2)}{3!} y^3 \ldots \qquad (25)$$

(**25**) is valid for all nonnegative (not necessarily integer) values of $g$ and for $|y| < 1$.

One possible approach of approximating $\left(\alpha(x)\right)^g$ is then given by :

$$\left(\alpha(x)\right)^g = \left(\theta + (\alpha-\theta)\, e^{-\lambda x}\right)^g = \theta^g \left(1 + \frac{(\alpha-\theta)}{\theta}\, e^{-\lambda x}\right)^g = \theta^g (1+y)^g$$

with $y \equiv \frac{(\alpha-\theta)}{\theta}\, e^{-\lambda x}$,

leading to:

$$\left(\alpha(x)\right)^g \approx \theta^g \Big(1 + g\theta^{-1}(\alpha-\theta)e^{-\lambda x} + \frac{1}{2} g\,(g-1)\theta^{-2}(\alpha-\theta)^2\, e^{-2\lambda x}$$
$$+ \frac{1}{6} g\,(g-1)(g-2)\theta^{-3}(\alpha-\theta)^3\, e^{-3\lambda x} \ldots\Big)$$

Note that, unlike $\left(\alpha(x)\right)^g$, the approximation is linear in exponential terms $e^{-\lambda x}, e^{-2\lambda x}, e^{-3\lambda x}, \ldots$, which can be integrated easily with respect to $x$.

Unfortunately, the validity condition $|y| < 1$ will be always satisfied only if $\alpha < 2\theta$, and this latter condition will be violated quite often in practice, resulting in erroneous approximations, in particular for short expiries and high instantaneous volatility[8]. To make the expansion a workable solution for all plausible parameter values, we need to reparametrize $\alpha(x)$ as follows:

$$\alpha(x) = \theta + (\alpha-\theta)\, e^{-\lambda x} = \mu + \mu\delta\,\left(2e^{-\lambda x} - 1\right)$$

with:

$$\mu = \frac{\alpha+\theta}{2} \quad \text{and} \quad \delta = \frac{\alpha-\theta}{\alpha+\theta}$$

---

[8] In particular, if $\alpha > 2\theta$ the condition $|y| < 1$ will be violated for all time points $x < \frac{1}{\lambda} ln \frac{(\alpha-\theta)}{\theta}$. Another simple binomial expansion for $\left(\alpha(x)\right)^g$, with factoring out $\alpha$ instead of $\theta$, leads to the analogous validity condition $\theta < 2\alpha$, which will also be violated above a certain $x$ in situations of low instantaneous volatility.

Then, $\big(\alpha(x)\big)^g$ becomes:

$$\big(\alpha(x)\big)^g = \mu^g \left(1 + \delta\left(2e^{-\lambda x} - 1\right)\right)^g = \mu^g (1+y)^g$$

with $y \equiv \delta\left(2e^{-\lambda x} - 1\right) = \frac{\alpha-\theta}{\alpha+\theta}\left(2e^{-\lambda x} - 1\right)$. With this reparameterization, the condition $|y| < 1$ is always satisfied for all plausible parameter values[9], and thus the binomial series expansion can be applied reliably, without parameter restrictions.

In this paper, we employ the expansion using the first 4 terms, resulting in[10]:

$$\big(\alpha(x)\big)^g = \left(\theta + (\alpha-\theta)\, e^{-\lambda x}\right)^g = \mu^g \left(1 + \delta\left(2e^{-\lambda x} - 1\right)\right)^g \approx \tag{26}$$

$$\approx \mu^g \Big(1 + g\delta\left(2e^{-\lambda x} - 1\right) + \frac{1}{2} g(g-1)\delta^2\left(2e^{-\lambda x} - 1\right)^2 + \frac{1}{6} g(g-1)(g-2)\delta^3\left(2e^{-\lambda x} - 1\right)^3\Big)$$

with $\mu = \frac{\alpha+\theta}{2}$ and $\delta = \frac{\alpha-\theta}{\alpha+\theta}$ as before.

Note that, after expanding the powers of $\left(2e^{-\lambda x} - 1\right)$, the approximation (**26**) is still linear in exponential terms $e^{-\lambda x}, \dots, e^{-3\lambda x}$, and thus can be integrated in closed form with respect to $x$.

Technically, we retorted again to *sympy* to approximate the problematic terms[11] from (**19**), and as before, for subsequent symbolic integration. Note that this approach does work for $\gamma$ specified as a parameter, as for $g$ in (**26**) above, but results, beyond the expansion order of 2, in very lengthy expressions upon integration. However, if $\gamma$ is preset as a constant, the length of the resulting expressions is comparable to that for *mrSABR* (where no approximative expansions were used). Appendix C to this paper contains the resulting closed-form expressions with $\gamma$ preset to 0.5 (*CIR-ZABR* model), with the binomial expansion order of 4, as specified in (**26**) above. The expressions are still simple ratios of multivariate polynomials.

---

[9] Note that: $0 \le e^{-\lambda x} < 1$ (for $x > 0$ expressing time), and $-1 < \delta = \frac{\alpha-\theta}{\alpha+\theta} < 1$ (for $\alpha > 0$, $\theta > 0$).

[10] We found that the expansions resulted in acceptable precision levels only starting with this expansion order. In principle, a higher order can be used analogously, but the resulting expressions get increasingly lengthy.

[11] We used the Taylor-series expansion routine of *sympy*. Note that the binomial expansion of $(1+y)^g$ is equivalent to (and is a special case of) classical Taylor-series expansion for $f(\mathrm{y}) = (1+y)^g$ with respect to y.

Note that for the special case $\alpha = \theta = \sigma$, the expressions in Appendix A and B reduce for *mrSABR* and *hSABR* to the closed-form formulas **(22)** and **(23)** already derived in the Hagan papers. Moreover, in that special case, the *CIR-ZABR* expressions in Appendix C also reduce to **(24)** when $\gamma = 0.5$, as the approximations applied are actually exact if $\alpha = \theta$.

The closed-form expressions for the three models in Appendix are our main theoretical result. Drawing on these expressions, we elaborate in the following section on the practical usage of the *hSABR*, *mrSABR* and *CIR-ZABR* models for calibrations to equity market data.

## 9. Calibration to equity volatility surfaces

We calibrated the five parameters $\alpha, \theta, \lambda, \rho, \nu$ of the three inspected models to market-implied volatility surfaces. To this end, the model predictions were calculated via:

- the expressions from Appendix A, along with the formulas in (**11**) for *hSABR*
- the expressions from Appendix B, along with the formulas in (**17**) for *mrSABR*
- the expressions from Appendix C, along with the formulas in (**17**) for *CIR-ZABR*

and subsequent application of the original SABR approximation (**7**) for all three models.

As the calibration target, we used historical volatility surfaces of EuroStoxx index options (as extracted from Bloomberg) at monthly intervals from 2021 to 2024, with 15 implied volatility quotes per surface (5 strikes × 3 maturities). For the technical parameter estimation, we employed the standard Excel Solver, minimizing the root mean squared error (RMSE) between the market quotes and the model predictions.

First, we performed the calibration of all five parameters for each observation date/surface $t$. The calibration was numerically unproblematic in all cases, yielding plausible time-dependent estimates $\hat{\alpha}_t, \hat{\theta}_t, \hat{\lambda}_t, \hat{\nu}_t, \hat{\rho}_t$, as depicted against $t$ in Figure 1 below. The resulting fit was strong across all three models for all surfaces. The average (or, alternatively, maximum) RMSE was only 0.8% (1.0%) for *mrSABR*, 0.7% (1.0%) for *hSABR*, and 0.7% (0.9%) for *CIR-ZABR*, measured in units of implied volatility[12]. The models explained approximately 99% of the original variance in implied volatilities across the 15 data points, while using only five parameters per surface.

[12] When the constant-expected-volatility restriction $\hat{\alpha}_t = \hat{\theta}_t$ was applied during estimation, the RMSE increased by about 0.5% on average (and by up to 5% for some surfaces), confirming the inadequacy of this restriction for equity markets.

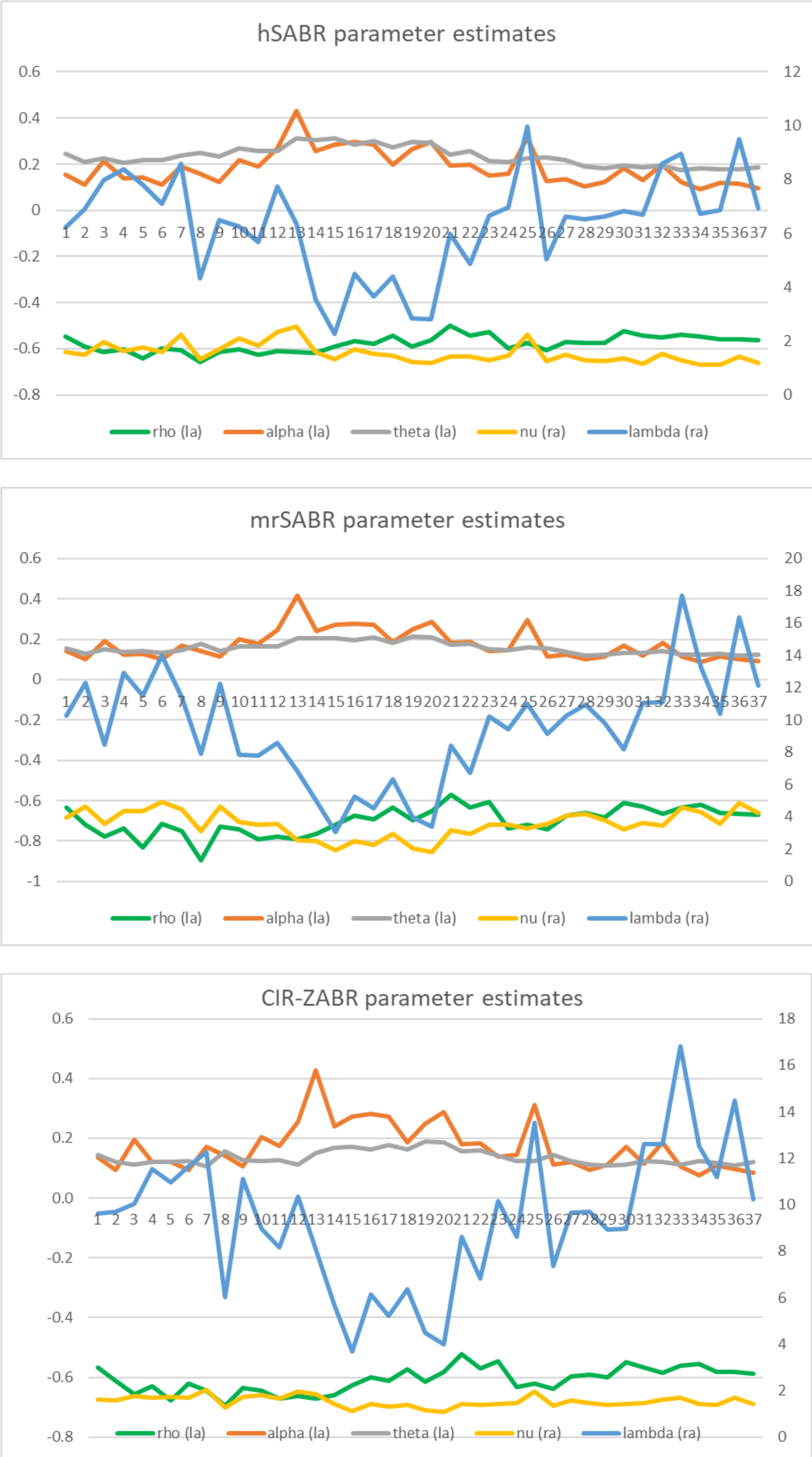

**Figure 3: Parameter estimates for the mean-reverting models (ra=right axis, la=left axis)**

A closer inspection of the results revealed some issues, though. Table 1 below reports the descriptive statistics and Spearman correlations for the parameter estimates.

***hSABR:***

| | *lambda* | *rho* | *alpha* | *theta* | *nu* |
|---|---|---|---|---|---|
| **Min** | 2.26 | -0.66 | 0.09 | 0.17 | 1.12 |
| **Median** | 6.64 | -0.57 | 0.16 | 0.23 | 1.47 |
| **Max** | 9.96 | -0.50 | 0.43 | 0.31 | 2.54 |
| **StdDev / Avg** | 29.13% | 5.97% | 40.87% | 17.82% | 22.90% |
| | | | | | |
| **Correlations:** | | *rho* | *alpha* | *theta* | *nu* |
| | *alpha* | -0.18 | | | |
| | *theta* | -0.40 | 0.79 | | |
| | *nu* | -0.54 | 0.45 | 0.42 | |
| | *lambda* | 0.05 | -0.39 | -0.71 | 0.24 |

***mrSABR:***

| | *lambda* | *rho* | *alpha* | *theta* | *nu* |
|---|---|---|---|---|---|
| **Min** | 3.07 | -0.90 | 0.08 | 0.12 | 1.83 |
| **Median** | 9.82 | -0.69 | 0.15 | 0.15 | 3.58 |
| **Max** | 17.69 | -0.57 | 0.42 | 0.21 | 4.92 |
| **StdDev / Avg** | 37.02% | 9.75% | 42.70% | 18.63% | 23.53% |
| | | | | | |
| **Correlations:** | | *rho* | *alpha* | *theta* | *nu* |
| | *alpha* | -0.26 | | | |
| | *theta* | -0.27 | 0.89 | | |
| | *nu* | 0.00 | -0.83 | -0.85 | |
| | *lambda* | 0.13 | -0.79 | -0.86 | 0.91 |

***CIR-ZABR:***

| | *lambda* | *rho* | *alpha* | *theta* | *nu* |
|---|---|---|---|---|---|
| **Min** | 3.69 | -0.70 | 0.07 | 0.11 | 1.09 |
| **Median** | 9.67 | -0.61 | 0.14 | 0.12 | 1.46 |
| **Max** | 16.82 | -0.52 | 0.43 | 0.19 | 2.04 |
| **StdDev / Avg** | 31.45% | 6.81% | 45.50% | 17.39% | 14.86% |
| | | | | | |
| **Correlations:** | | *rho* | *alpha* | *theta* | *nu* |
| | *alpha* | -0.28 | | | |
| | *theta* | 0.04 | 0.48 | | |
| | *nu* | -0.34 | -0.05 | -0.57 | |
| | *lambda* | 0.19 | -0.54 | -0.73 | 0.66 |

**Table 1: Descriptive statistics and correlation for the parameter estimates (across time)**

First, in all three models, the estimates $\hat{\theta}_t$ (long-term volatility) correlate positively with the estimates $\hat{\alpha}_t$ (initial volatility). Also, the estimates $\hat{\lambda}_t$ (mean-reverting speed) correlate negatively to both $\hat{\theta}_t$ and $\hat{\alpha}_t$. This might indicate a rather non-linear mean-reverting pattern implied in the market data: during the periods of strongly above-average instantaneous volatility, the market seems to project a reversion to a higher long-term volatility and at a slower (linearly specified) speed. The positive correlation between the estimates $\hat{\lambda}_t$ (mean-reverting speed) and $\hat{\nu}_t$ (volatility of volatility) is, to a certain extent, theoretically expected, as both parameters influence the convexity (deepness) of the smile[13].

Finally, the correlation between the estimates $\hat{\alpha}_t$ (initial volatility) and $\hat{\nu}_t$ (volatility of volatility) is quite outstanding: it is strongly negative in the *mrSABR* case, moderately positive in the *hSABR* case, and close to 0 in the *CIR-ZABR* case. Most probably, this is due to the true volatility process being closer to CIR diffusion as specified in *CIR-ZABR*. Remember that the volatility is modelled with lognormal diffusion in *mrSABR*, and with approximately normal diffusion in *hSABR*. The correlations observed would be actually expected if the true volatility diffusion were between normal and lognormal (e.g. of the CIR-type)[14].

The above correlations in parameter estimates are generally benign for the purposes of interpolation and/or extrapolation of the surfaces. However, they need to be accounted for if some historically estimated parameters should be applied to new market data. Note that the only formally time-dependent (stochastic) factors in the analyzed mean-reverting models are the forward price and the instantaneous volatility. Thus, $f$ and $\alpha$ need to be determined anew for a new market situation (or simulated e.g. via increments of their last known values). On the opposite, the parameters $\theta, \lambda, \nu, \rho$ are formally specified as constant (non-stochastic). However, because of the correlations mentioned above, if these parameters are applied unchanged (e.g. estimated from old data) along with a new $\alpha$, the generated smiles / prices might be biased, especially if the correlations between the estimates for $\alpha$ vs $\theta, \lambda, \nu, \rho$ are significant. In this sense, the CIR-ZABR model appears least problematic[15], with only moderate or vanishing correlations to $\alpha$.

---

[13] E.g., a higher estimate for $\nu$ can be offset by a higher estimate for $\lambda$, resulting in a similar smile convexity (as already reported for the original Heston model e.g. in *Cui et al. (2017)*). Thus, this positive correlation seems to reflect the (moderate) redundancy immanently present in the mean-reverting models. Continuing the example, the higher resulting estimate for $\lambda$ would also influence, apart from the smile convexity, the term structure of volatilities. And this latter effect can itself be offset by a parameter $\theta$ being estimated closer to $\alpha$. This might contribute to the correlations between $\alpha_t, \theta_t, \lambda_t$ as described above.

[14] In particular, in that case the *hSABR* model would be expected to estimate a higher $\nu_t$ for those surfaces where the initial instantaneous volatility $\alpha_t$ happens to be high (and vice versa), in order to compensate for the misspecification. For the *mrSABR* model, the opposite would be the case.

[15] The correlation between $\nu$ and $\alpha$ is of particular concern for equity markets. E.g., in a stress scenario of significant negative returns and sharply rising volatility $\alpha$, and if some old (or average) $\nu$ is used, *mrSABR* would

Another way to verify the consistency of parameter estimates is to inspect the prediction errors when the models are re-estimated per each observation date, with a different $\hat{\alpha}_t$ per observation date $t$ as before, but when some (or all) of the other parameters $\theta_t, \lambda_t, \nu_t, \rho_t$ are instead either fixed at the level of their previous-period estimates $\hat{\theta}_{t-1}, \hat{\lambda}_{t-1}, \hat{\nu}_{t-1}, \hat{\rho}_{t-1}$, or estimated as overall (time-independent) estimates $\hat{\theta}, \hat{\lambda}, \hat{\nu}, \hat{\rho}$. Table 2 below reports the results.

| time-independent parameters, except: | hSABR | mrSABR | CIR-ZABR |
|---|---|---|---|
| alpha | 1.4% | 1.5% | 1.5% |
| alpha = theta | 1.6% | 1.4% | 1.5% |
| alpha & theta | 1.2% | 1.1% | 1.1% |
| alpha & lambda | 1.3% | 1.4% | 1.1% |
| alpha & nu | 1.3% | 1.4% | 1.2% |
| alpha & pho | 1.4% | 1.4% | 1.3% |

| previous-period parameters, except: | hSABR | mrSABR | CIR-ZABR |
|---|---|---|---|
| alpha | 1.2% | 1.2% | 1.1% |
| alpha = theta | 1.6% | 1.3% | 1.4% |
| alpha & theta | 1.1% | 1.0% | 1.0% |
| alpha & lambda | 1.1% | 1.1% | 1.0% |
| alpha & nu | 1.0% | 1.1% | 0.9% |
| alpha & pho | 1.0% | 1.1% | 1.0% |

**Table 2: Root mean squared error, time-independent and previous-period parameters**

In particular, when all coefficients, except the explicitly stochastic $\hat{\alpha}$, are estimated as time-independent (or, alternatively, previous-period) parameters, the RMSE increases quite materially from the above mentioned 0.7-0.8%: to 1.4-1.5% (or 1.1-1,2%). This might advocate accounting for other parameters, apart from $\hat{\alpha}_t$, as stochastic factors: with two parameters re-estimated per surface, the RMSE improves by up to 0.4%. Overall, the re-estimation of long-term volatility $\theta_t$ seems to improve the RMSE most.

Lastly, we note that for *hSABR*, the Feller non-degeneracy condition (**3**) was significantly violated in all inspected surfaces. Conversely, for *mrSABR*, the non-degeneracy condition (**13**) was satisfied in all estimates. For *CIR-ZABR*, the corresponding condition (**21**) was satisfied in most, but not all, cases, with only mild violations.

estimate a smile that is too deep, while *hSABR* would estimate one that is too flat. Of course, the correlation can be accounted for using workarounds, such as regressing $\hat{\nu}_t$ on $\hat{\alpha}_t$ and using the regression predictions in the scenario. However, it is preferable for the model itself to account for the correlation patterns.

## 10. Conclusions and areas of application

In this paper, we have shown that entire equity volatility surfaces can be easily modeled using SABR-based mean-reverting-volatility models with only five parameters, via closed-form expressions for implied volatilities, as derived above.

These SABR-based models provide a viable, simple, and fast alternative to modeling surfaces using the classical Heston model or more complex local/stochastic volatility models. They may be particularly advantageous in unsupervised, automated, or computationally demanding settings (e.g., Monte Carlo simulations) or in scenarios where advanced numerical integration is problematic (such as in Excel-based tools).

Empirically, all three investigated SABR-based models produced similar, high-quality fits to the equity volatility surfaces across all observation dates. However, the three models—and likely the original Heston model as well—yield high correlations among some parameter estimates over time. In particular, a material correlation between the parameter estimates for instantaneous volatility and volatility-of-volatility is observed in both the mean-reverting SABR (*mrSABR*) and approximated Heston (*hSABR*) models, but not in the mean-reverting ZABR (*CIR-ZABR*) model. This suggests that the true volatility process for equities is more accurately described by a CIR diffusion rather than the lognormal diffusion assumed in SABR or the normal diffusion implicitly assumed in Heston. This might have important implications for risk management.

Furthermore, while the *hSABR* model results in shorter closed-form expressions, it often violated the non-degeneracy Feller condition. This makes it better suited for surface interpolation, or for valuations based on recently estimated parameters, or in situations where the classical Heston model should be preferred. For the purposes of extrapolating and simulating volatility surfaces, the *CIR-ZABR* model appears to be the preferable solution. Its parameter estimates exhibit only moderate correlations and typically satisfy non-degeneracy conditions, making this model potentially most reliable for these purposes. As for *mrSABR*, despite its excellent non-degenerative fit, it exhibits high parameter correlations. This makes it suitable for interpolation and extrapolation, but less so for estimating or simulating new volatility surfaces.

Finally, the parameter estimates from all three models suggest nonlinear mean reversion in instantaneous volatility (or, alternatively, a stochastic long-term volatility) in equity markets, warranting further research.

## Appendix A: Closed-form expressions for *hSABR*

The following formulas for *hSABR* are in Excel format and use notations as follows:
`alps`: parameter $\alpha^2$ (initial instantaneous variance)
`thes`: parameter $\theta^2$ (long-term variance)
`lam`: parameter $\lambda$ (speed of mean reversion of variance)
`nu`: parameter $\nu$ (CIR volatility of variance)
`rho`: parameter $\rho$ (correlation)
`T_ex`: maturity/expiry of option
`E`: exponential term calculated as `EXP(lam*T_ex)`

### Formula for $\boldsymbol{\tau_{ex}}$ in *hSABR*

```
(T_ex*lam*thes*E - alps + thes + E*(alps - thes))
/
(lam*E)
```

### Formula for $\bar{\boldsymbol{b}}$ in *hSABR*

```
nu*rho*E*(T_ex*lam*( - 1*alps + thes*E + thes) + alps*E - alps - 2*thes*E + 2*thes)
/
(T_ex*lam*thes*E - alps + thes + E*(alps - thes))^2
```

### Formula for $\bar{\boldsymbol{c}}$ in *hSABR*

```
- 3*nu^2*E*(8*rho^2*E*(T_ex*lam*( - 1*alps + thes*E + thes) + alps*E - alps - 2*thes*E + 2*thes)^2 -
(T_ex*lam*thes*E - alps + thes + E*(alps - thes))*(2*T_ex*lam*thes*E^2*(4*rho^2 + 1) - 2*alps +
8*rho^2*E^2*(alps - 3*thes) - 4*rho^2*E*(T_ex^2*alps*lam^2 - T_ex^2*lam^2*thes + 2*T_ex*alps*lam -
4*T_ex*lam*thes + 2*alps - 6*thes) + thes + E^2*(2*alps - 5*thes) + 4*E*( - 1*T_ex*alps*lam + T_ex*lam*thes
+ thes)))
/
(8*(T_ex*lam*thes*E - alps + thes + E*(alps - thes))^4)
```

## Appendix B: Closed-form expressions for *mrSABR*

The following formulas for *mrSABR* are in Excel format and use notations as follows:
`alp`: parameter $\alpha$ (initial instantaneous volatility)
`the`: parameter $\theta$ (long-term volatility)
`lam`: parameter $\lambda$ (speed of mean reversion of volatility)
`nu`: parameter $\nu$ (lognormal volatility of volatility)
`rho`: parameter $\rho$ (correlation)
`T_ex`: maturity/expiry of option
`E`: exponential term calculated as `EXP(lam*T_ex)`

### Formula for $\boldsymbol{\tau_{ex}}$ in *mrSABR*

```
(2*T_ex*lam*the^2*E^2 - alp^2 + 2*alp*the - the^2 - 4*the*E*(alp - the) + E^2*(alp^2 + 2*alp*the -
3*the^2))
/
(2*lam*E^2)
```

### Formula for $\boldsymbol{G_{int}}$ in *mrSABR*

```
nu^2*( - 2*T_ex*alp^2*lam + 4*T_ex*alp*lam*the + 2*T_ex*lam*the^2*E^2 - 2*T_ex*lam*the^2 - alp^2 +
6*alp*the - 4*the^2 - 8*the*E*(alp - the) - E^2*( - 1*alp^2 - 2*alp*the + 4*the^2))
/
(4*lam^2*E^2)
```

### Formula for $\bar{\boldsymbol{b}}$ in *mrSABR*

```
- 4*nu*rho*E*(6*T_ex*alp^2*lam*the*E + 12*T_ex*alp*lam*the^2*E^2 - 12*T_ex*alp*lam*the^2*E -
6*T_ex*lam*the^3*E^3 - 12*T_ex*lam*the^3*E^2 + 6*T_ex*lam*the^3*E - alp^3*E^3 + 3*alp^3*E - 2*alp^3 -
3*alp^2*the*E^3 + 6*alp^2*the*E^2 - 9*alp^2*the*E + 6*alp^2*the - 6*alp*the^2*E^3 + 6*alp*the^2*E^2 +
6*alp*the^2*E - 6*alp*the^2 + 16*the^3*E^3 - 18*the^3*E^2 + 2*the^3)
/
(3*(2*T_ex*lam*the^2*E^2 - alp^2 + 2*alp*the - the^2 - 4*the*E*(alp - the) + E^2*(alp^2 + 2*alp*the -
3*the^2))^2)
```

### Formula for $\bar{\boldsymbol{c}}$ in *mrSABR*

```
nu^2*E^2*( - 32*rho^2*(6*T_ex*alp^2*lam*the*E + 12*T_ex*alp*lam*the^2*E^2 - 12*T_ex*alp*lam*the^2*E -
6*T_ex*lam*the^3*E^3 - 12*T_ex*lam*the^3*E^2 + 6*T_ex*lam*the^3*E - alp^3*E^3 + 3*alp^3*E - 2*alp^3 -
3*alp^2*the*E^3 + 6*alp^2*the*E^2 - 9*alp^2*the*E + 6*alp^2*the - 6*alp*the^2*E^3 + 6*alp*the^2*E^2 +
6*alp*the^2*E - 6*alp*the^2 + 16*the^3*E^3 - 18*the^3*E^2 + 2*the^3)^2 + 9*(2*T_ex*lam*the^2*E^2 - alp^2 +
2*alp*the - the^2 - 4*the*E*(alp - the) + E^2*(alp^2 + 2*alp*the - 3*the^2))*(4*T_ex*alp^4*lam -
16*T_ex*alp^3*lam*the + 24*T_ex*alp^2*lam*the^2 - 16*T_ex*alp*lam*the^3 + 16*T_ex*lam*the^4*E^4*(5*rho^2 +
1) + 4*T_ex*lam*the^4 - 12*alp^4*rho^2 + 3*alp^4 + 48*alp^3*rho^2*the - 20*alp^3*the - 72*alp^2*rho^2*the^2
+ 40*alp^2*the^2 + 48*alp*rho^2*the^3 - 32*alp*the^3 - 12*rho^2*the^4 + 9*the^4 -
8*the*E^3*(8*T_ex^2*alp*lam^2*rho^2*the^2 - 8*T_ex^2*lam^2*rho^2*the^3 + 8*T_ex*alp^2*lam*rho^2*the +
16*T_ex*alp*lam*rho^2*the^2 + 10*T_ex*alp*lam*the^2 - 32*T_ex*lam*rho^2*the^3 - 10*T_ex*lam*the^3 +
4*alp^3*rho^2 + alp^3 + 8*alp^2*rho^2*the + 5*alp^2*the + 4*alp*rho^2*the^2 - 10*alp*the^2 -
36*rho^2*the^3) + E^4*(4*alp^4*rho^2 + alp^4 + 16*alp^3*rho^2*the + 4*alp^3*the + 40*alp^2*rho^2*the^2 +
8*alp^2*the^2 + 80*alp*rho^2*the^3 + 16*alp*the^3 - 292*rho^2*the^4 - 53*the^4) + 4*E^2*( -
16*T_ex^2*alp^2*lam^2*rho^2*the^2 + 32*T_ex^2*alp*lam^2*rho^2*the^3 - 16*T_ex^2*lam^2*rho^2*the^4 -
16*T_ex*alp^3*lam*rho^2*the + 40*T_ex*alp^2*lam*rho^2*the^2 - 6*T_ex*alp^2*lam*the^2 -
16*T_ex*alp*lam*rho^2*the^3 + 12*T_ex*alp*lam*the^3 - 8*T_ex*lam*rho^2*the^4 - 6*T_ex*lam*the^4 -
6*alp^4*rho^2 - alp^4 + 16*alp^3*rho^2*the - 4*alp^3*the + 8*alp^2*rho^2*the^2 + 36*alp^2*the^2 -
32*alp*rho^2*the^3 - 60*alp*the^3 + 12*rho^2*the^4 + 27*the^4) + 8*E*(8*T_ex*alp^3*lam*rho^2*the +
2*T_ex*alp^3*lam*the - 24*T_ex*alp^2*lam*rho^2*the^2 - 6*T_ex*alp^2*lam*the^2 + 24*T_ex*alp*lam*rho^2*the^3
+ 6*T_ex*alp*lam*the^3 - 8*T_ex*lam*rho^2*the^4 - 2*T_ex*lam*the^4 + 4*alp^4*rho^2 - 12*alp^3*rho^2*the +
5*alp^3*the + 8*alp^2*rho^2*the^2 - 19*alp^2*the^2 + 4*alp*rho^2*the^3 + 22*alp*the^3 - 4*rho^2*the^4 -
8*the^4)))
/
(6*(2*T_ex*lam*the^2*E^2 - alp^2 + 2*alp*the - the^2 - 4*the*E*(alp - the) + E^2*(alp^2 + 2*alp*the -
3*the^2))^4)
```

## Appendix C: Closed-form expressions for *CIR-ZABR*

*CIR-ZABR* is a special case of *mrZABR*, with parameter $\gamma$ (CEV parameter in volatility process) fixed at ½. The results below stem from approximations via fourth-order binomial expansions.

The formulas are in Excel format and use notations as follows:
`alp`: parameter $\alpha$ (initial instantaneous volatility)
`the`: parameter $\theta$ (long-term volatility)
with the formulas below using the **reparameterization**:
`mu=(alp + the)/2` and `dlt=(alp - the)/(alp + the)`
so that: `alp = mu*(1 + dlt)` and `the = mu*(1 − dlt)`

`lam`: parameter $\lambda$ (speed of mean reversion of volatility)
`nu`: parameter $\nu$ (CIR-volatility of volatility)
`rho`: parameter $\rho$ (correlation)
`T_ex`: maturity/expiry of option
`E`: exponential term calculated as `EXP(lam*T_ex)`

### Formula for $\boldsymbol{\tau_{ex}}$ in *CIR-ZABR*

```
mu^2*(T_ex*dlt^2*lam*E^2 - 2*T_ex*dlt*lam*E^2 + T_ex*lam*E^2 - 2*dlt^2*E^2 + 4*dlt^2*E - 2*dlt^2 +
4*dlt*E^2 - 4*dlt*E)
/  (lam*E^2)
```

### Formula for $\boldsymbol{G_{int}}$ in *CIR-ZABR*

```
mu*nu^2*(-2*T_ex*lam*E^2*(dlt - 1) - 8*dlt*E + 3*dlt + E^2*(5*dlt - 1) + 1)
/  (4*lam^2*E^2)
```

### Formula for $\boldsymbol{\bar{b}}$ in *CIR-ZABR*

```
- 1*nu*rho*E*( - 72*T_ex*dlt^2*lam*E*(dlt - 2) + 18*T_ex*dlt*lam*E^2*(3*dlt^2 - 12*dlt + 8) +
3*T_ex*lam*E^3*(5*dlt^3 - 30*dlt^2 + 40*dlt - 16) - 44*dlt^3 + 12*dlt*E*(3*dlt^2 + 9*dlt - 4) -
E^3*(49*dlt^3 - 270*dlt^2 + 312*dlt - 48) + E^2*(57*dlt^3 - 378*dlt^2 + 360*dlt - 48))
/
(6*mu^(3/2)*( - 2*T_ex*lam*E^2*(dlt - 1)^2 - 8*dlt*E*(dlt - 1) + E^2*(3*(dlt - 1)^2 + 2*(dlt - 1)*(dlt + 1)
- (dlt + 1)^2) + (dlt - 1)^2 + 2*(dlt - 1)*(dlt + 1) + (dlt + 1)^2)^2)
```

### Formula for $\boldsymbol{\bar{c}}$ in CIR-ZABR:

$$\bar{\boldsymbol{c}} = 3\boldsymbol{c_{int}}/\tau_{ex}^3 - 3\bar{\boldsymbol{b}}^{\boldsymbol{2}}$$

with $\boldsymbol{c_{int}}$ as follows:

```
mu^3*nu^2*(30*T_ex*lam*E^6*(dlt^6*rho^2 + 12*dlt^5*rho^2 - 12*dlt^4*rho^2 - 1024*dlt^3*rho^2 - 256*dlt^3 +
3072*dlt^2*rho^2 + 768*dlt^2 - 3072*dlt*rho^2 - 768*dlt + 1024*rho^2 + 256) - 80*dlt^6*rho^2 +
192*dlt^4*rho^2*E*(3*dlt^2 + 6*dlt - 8) - 30*dlt^2*E^2*( - 12*T_ex*dlt^4*lam*rho^2 -
48*T_ex*dlt^3*lam*rho^2 + 384*T_ex*dlt^2*lam*rho^2 + 47*dlt^4*rho^2 + 168*dlt^3*rho^2 + 544*dlt^2*rho^2 -
96*dlt*rho^2 + 384*dlt + 128) + 160*dlt*E^3*( - 9*T_ex*dlt^5*lam*rho^2 - 54*T_ex*dlt^4*lam*rho^2 +
36*T_ex*dlt^3*lam*rho^2 + 396*T_ex*dlt^2*lam*rho^2 + 7*dlt^5*rho^2 + 27*dlt^4*rho^2 + 294*dlt^3*rho^2 +
316*dlt^2*rho^2 + 400*dlt^2 - 132*dlt*rho^2 - 96*dlt - 48) - E^6*(131*dlt^6*rho^2 + 1512*dlt^5*rho^2 -
2316*dlt^4*rho^2 - 125120*dlt^3*rho^2 - 28160*dlt^3 + 347520*dlt^2*rho^2 + 76800*dlt^2 - 299520*dlt*rho^2 -
65280*dlt + 57600*rho^2 + 11520) - 60*E^5*(576*T_ex^2*dlt^3*lam^2*rho^2 - 1152*T_ex^2*dlt^2*lam^2*rho^2 +
576*T_ex^2*dlt*lam^2*rho^2 - 6*T_ex*dlt^6*lam*rho^2 - 60*T_ex*dlt^5*lam*rho^2 + 144*T_ex*dlt^4*lam*rho^2 +
2016*T_ex*dlt^3*lam*rho^2 + 768*T_ex*dlt^3*lam - 5184*T_ex*dlt^2*lam*rho^2 - 1536*T_ex*dlt^2*lam +
3456*T_ex*dlt*lam*rho^2 + 768*T_ex*dlt*lam - 384*T_ex*lam*rho^2 + dlt^6*rho^2 - 116*dlt^4*rho^2 +
2336*dlt^3*rho^2 + 128*dlt^3 - 7008*dlt^2*rho^2 - 1024*dlt^2 + 5760*dlt*rho^2 + 1152*dlt - 1024*rho^2 -
256) - 15*E^4*(36*T_ex^2*dlt^6*lam^2*rho^2 + 288*T_ex^2*dlt^5*lam^2*rho^2 - 4608*T_ex^2*dlt^3*lam^2*rho^2 +
4608*T_ex^2*dlt^2*lam^2*rho^2 - 48*T_ex*dlt^6*lam*rho^2 - 264*T_ex*dlt^5*lam*rho^2 -
864*T_ex*dlt^4*lam*rho^2 - 2304*T_ex*dlt^3*lam*rho^2 - 3072*T_ex*dlt^3*lam + 9984*T_ex*dlt^2*lam*rho^2 +
3072*T_ex*dlt^2*lam - 3072*T_ex*dlt*lam*rho^2 + dlt^6*rho^2 - 72*dlt^5*rho^2 + 2564*dlt^4*rho^2 +
2560*dlt^3*rho^2 + 4864*dlt^3 + 3456*dlt^2*rho^2 - 2304*dlt^2 - 3072*dlt*rho^2 - 768*dlt + 256*rho^2 +
256))
/  (7680*lam^3*E^6)
```